\documentclass[twocolumn,twoside,slac_two]{revtex4}
\usepackage{graphicx}
\usepackage{fancyhdr}
\def\kbar{\overline{K}{}^{\,0}}
\def\dbar{\overline{D}{}^{\,0}}

\def\cp{$CP$}
\def\cpv{$CPV$}
\def\ra{\!\rightarrow\!}

\def\dklnu{$D^0\ra K^+\ell^-\nu$}
\def\dkpi{$D^0\ra K^+\pi^-$}
\def\dkk{$D^0\ra K^+K^-$}
\def\dpipi{$D^0\ra\pi^+\pi^-$}
\def\dkkpp{$D^0\ra K^+K^-/\pi^+\pi^-$}
\def\dkspp{$D^0\ra K^0_S\,\pi^+\pi^-$}
\def\ycp{$y^{}_{\rm CP}$}

\def\gevm{~GeV/$c^2$}

\def\geve{~GeV}
\def\mevm{~MeV/$c^2$}
\def\meve{~MeV}

\def\babar{{\it BABAR\/}}

\def\simge{\mathrel{%
   \rlap{\raise 0.511ex \hbox{$>$}}{\lower 0.511ex \hbox{$\sim$}}}}
\def\simle{\mathrel{
   \rlap{\raise 0.511ex \hbox{$<$}}{\lower 0.511ex \hbox{$\sim$}}}}

\begin{document}

\title{\boldmath Searches for Mixing and \cp\ Violation in the
$D^0$-$\dbar$ System}

\author{A. J. Schwartz}
\affiliation{Department of Physics, University of Cincinnati,
P.O. Box 210011, Cincinnati, Ohio 45221, USA}

\begin{abstract}
We review searches for mixing and \cp\ violation in the
$D^0$-$\dbar$ system and discuss the first evidence for 
mixing recently obtained by the Belle and \babar\ collaborations. 
We also present world average values for the mixing parameters
$x=\Delta m/\Gamma$ and $y=\Delta\Gamma/(2\Gamma)$ as 
calculated by the Heavy Flavors Averaging Group.
\end{abstract}

\maketitle

\thispagestyle{fancy}

\section{Introduction}

Mixing in the $D^0$-$\dbar$ system has been searched for for more than 
two decades without success --- until this past year. Both ``$B$-factory''
experiments, Belle and \babar, have recently published evidence for
this phenomenon~\cite{belle_kk,babar_kpi}. Here we review these 
measurements and discuss their implications. In total four 
measurements are presented, involving the following decay 
modes~\cite{charge_conjugates}: \dklnu, \dkkpp, \dkpi, and \dkspp.

Mixing in heavy flavor systems such as that of $B^0$ and $B^0_s$ 
is governed by the short-distance box diagram. However, in the 
$D^0$ system this diagram is both doubly-Cabibbo-suppressed and
GIM-suppressed relative to the amplitude dominating the decay
width, and thus the short-distance rate is very small. Consequently, 
$D^0$-$\dbar$ mixing is expected to be dominated by long-distance 
processes that are difficult to calculate; theoretical estimates 
for the mixing parameters $x\equiv\Delta m/\Gamma$ and 
$y\equiv\Delta \Gamma/(2\Gamma)$ range over two-three orders 
of magnitude~\cite{Petrov}. Here, $\Delta M$ and $\Delta \Gamma$ 
are the differences in the masses and decay widths, respectively, 
of the two $D^0$-$\dbar$ mass eigenstates, and $\Gamma$ is the 
mean decay width.

All methods discussed here identify the flavor of the $D^0/\dbar$ 
when produced by reconstructing the decay $D^{*+}\ra D^0\pi^+$ or 
$D^{*-}\ra\dbar\pi^-$; the charge of the accompanying pion identifies 
the $D$ flavor. Because $M^{}_{D^*}-M^{}_{D^0}-M^{}_{\pi^+}\approx 6$\meve,
which is relatively small, the pion has very low momentum and thus is 
denoted $\pi^{}_s$ (``$\pi$ slow''). The $D^0$ decay time ($t$) 
is calculated via $(\ell/p)\times M^{}_D$, where $\ell$ is
the distance between the $D^*$ and $D^0$ decay vertices and 
$p$ is the $D^0$ momentum. The $D^*$ vertex position is taken to 
be the intersection of the $\pi^{}_s$ momentum with the beamspot 
profile. Most of the precision on $t$ is due to the vertical
($y$) component of $\ell$, as the spread of the beamspot is only 
a few microns in this dimension. To reject $D^{(*)}$ decays 
originating from $B$ decays, one requires $p^{}_{D^*}>2.5$\geve, 
which is the kinematic endpoint.

\section{\boldmath ``Wrong-sign'' \dklnu\ Decays}

The decay \dklnu\ is in principle an ideal signature 
for mixing, as this ``wrong-sign'' (WS) final state 
can be reached only via a $D^0\ra\dbar$ transition
(in contrast to the ``right-sign'' (RS) final state 
$K^-\ell^+\nu$). However, 
the neutrino in the final state precludes the decay from 
being fully-reconstructed, and consequently the decay time 
measurement is smeared. The decay time dependence is given by
$e^{-\Gamma t}[1+(x^2+y^2)t^2/2]$, and both Belle and \babar\ 
measure the coefficient $(x^2+y^2)/2\equiv R^{}_M$ by fitting the 
decay time distribution. To reduce backgrounds, only the 
electron channel ($Ke\nu$) has been used. 

Belle uses 253~fb$^{-1}$ of data~\cite{belle_semi}, making a 
relatively loose event selection and fitting the resulting 
$M(Ke\nu)$ and $\Delta M\equiv M(\pi^{}_s\,Ke\nu)-M(Ke\nu)$ 
distributions to determine the yield of WS events.
To improve the resolution on both $\Delta M$ and the 
decay time, two corrections are made to better determine
$P^{}_\nu = P^{}_{\rm cms} - P^{}_{\pi K e} - P^{}_{\rm rest}$,
where $P^{}_{\rm rest}$ is the total four-momentum of all
tracks and photons in the event besides the $\pi^{}_s$, $K$, and
$e$ candidates, and $P^{}_{\rm cms}$ is the four-momentum of
the center-of-mass (cms) system. First, $P^{}_{\rm rest}$ is 
rescaled by a factor $x$ such that 
$(P^{}_{\rm cms}-xP^{}_{\rm rest})^2=P^2_{D^*}=M^2_{D^*}$;
second, $\vec{p}^{}_{\rm rest}$ is rotated in the plane of
$\vec{p}^{}_{\rm rest}$ and $\vec{p}^{}_{\pi^{}_s Ke}$ such
that $(P^{}_{\rm cms}-P^{}_{\pi^{}_s Ke}-xP^{}_{\rm rest})^2=P^2_\nu=0$.

The ratio of WS to RS events are measured for six bins of decay 
time ranging from one to ten $D^0$ lifetimes. Summing over all 
bins, $16430\,\pm 155$ RS and $-1\,\pm 37$ WS events are found.
For each bin the ratio depends on $R^{}_M$, and the six resulting 
values of $R^{}_M$ are fit to a constant. The result is
\begin{eqnarray}
R^{}_M & = & (0.020\,\pm 0.047\,\pm 0.014)\%\,,
\end{eqnarray}
where the first error listed is statistical and the second
error is systematic. The latter is dominated by uncertainty in the
background $\Delta M$ PDF. As much of the $R^{}_M$ likelihood
is in the unphysical (negative) region, Belle uses a 
Feldman-Cousins~\cite{FeldmanCousins} approach to calculate
an upper limit; the result is $R^{}_M < 0.10\%$ at 90\% C.L. 
This limit implies both $|x|$ and  $|y|$ are $<4.5$\%.

The \babar\ experiment uses 344~fb$^{-1}$ of data~\cite{babar_semi} 
and, in contrast to Belle, imposes tight selection criteria to reduce 
background as much as possible (with a corresponding loss in efficiency). 
The most restrictive criterion is that events must have a $D^{(*)}$ 
decay fully reconstructed in the hemisphere opposite that of the 
semileptonic decay, where the modes $D^{*+}\ra D^0\pi^+$, 
$D^0\ra K^-\pi^+$, $D^0\ra K^-\pi^+\pi^0$, $D^0\ra K^-\pi^+\pi^+\pi^-$, 
and $D^+\ra K^-\pi^+\pi^+$ are used. This ``double-tagging'' 
eliminates copious background from real $D^0$'s combining 
with random $\pi^\pm$ tracks to make false $D^*$ 
candidates, but it reduces the signal efficiency by an 
order of magnitude. The neutrino momentum is determined 
via a neural network algorithm, and events are required 
to have decay times in the range 600-3900~fs (corresponding 
to 1.5-9.5 $D^0$ lifetimes). A separate neural network is 
used to select signal events, and a final set of kinematic
selection criteria are applied to the signal side. These
criteria include cuts on the $\pi^{}_s$ and $e$ momenta, 
and on the $dE/dx$ of the electron track in the silicon 
vertex tracker. 

\begin{figure}[h]
\centering
\includegraphics[width=40mm]{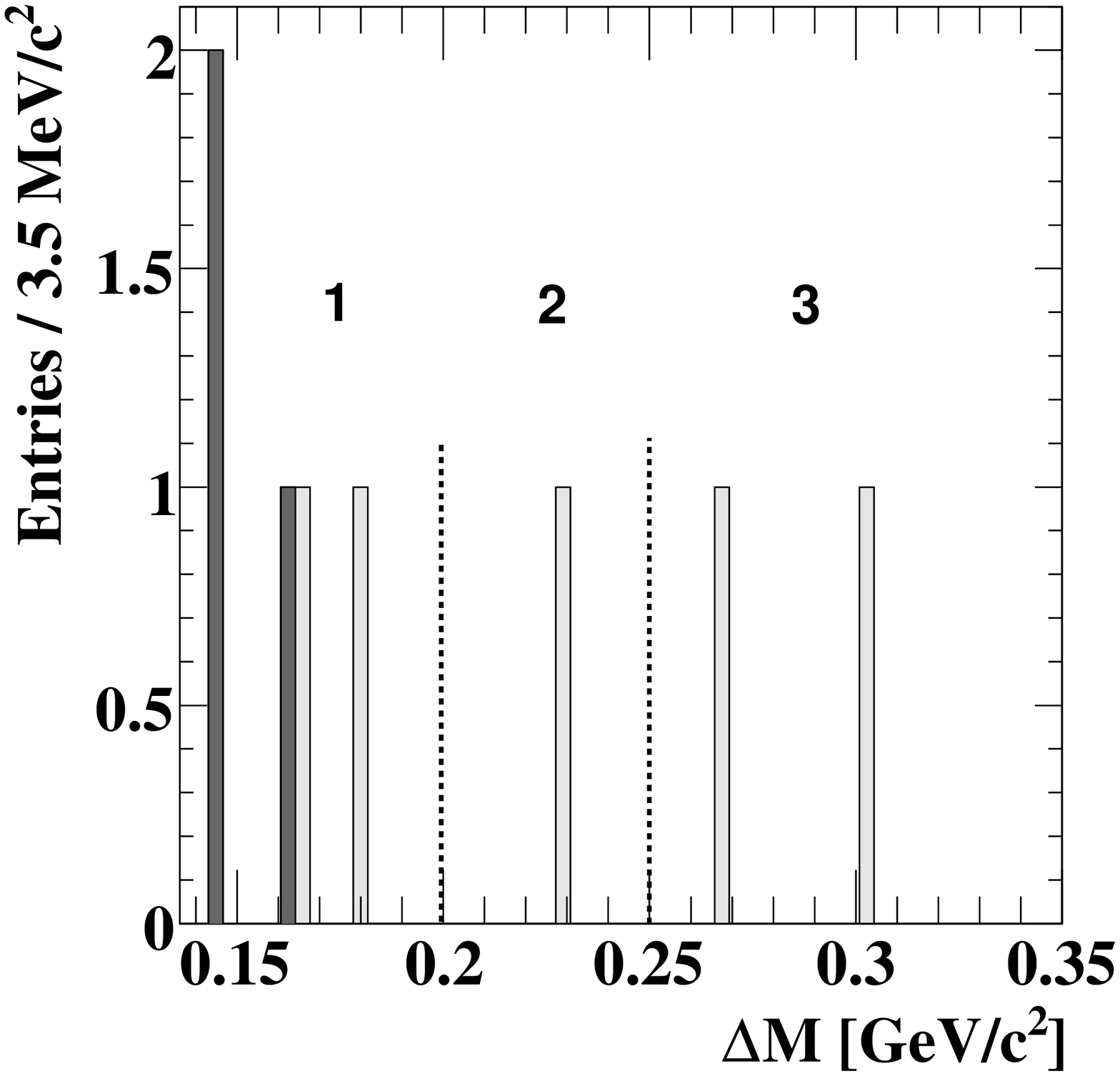}
\includegraphics[width=40mm]{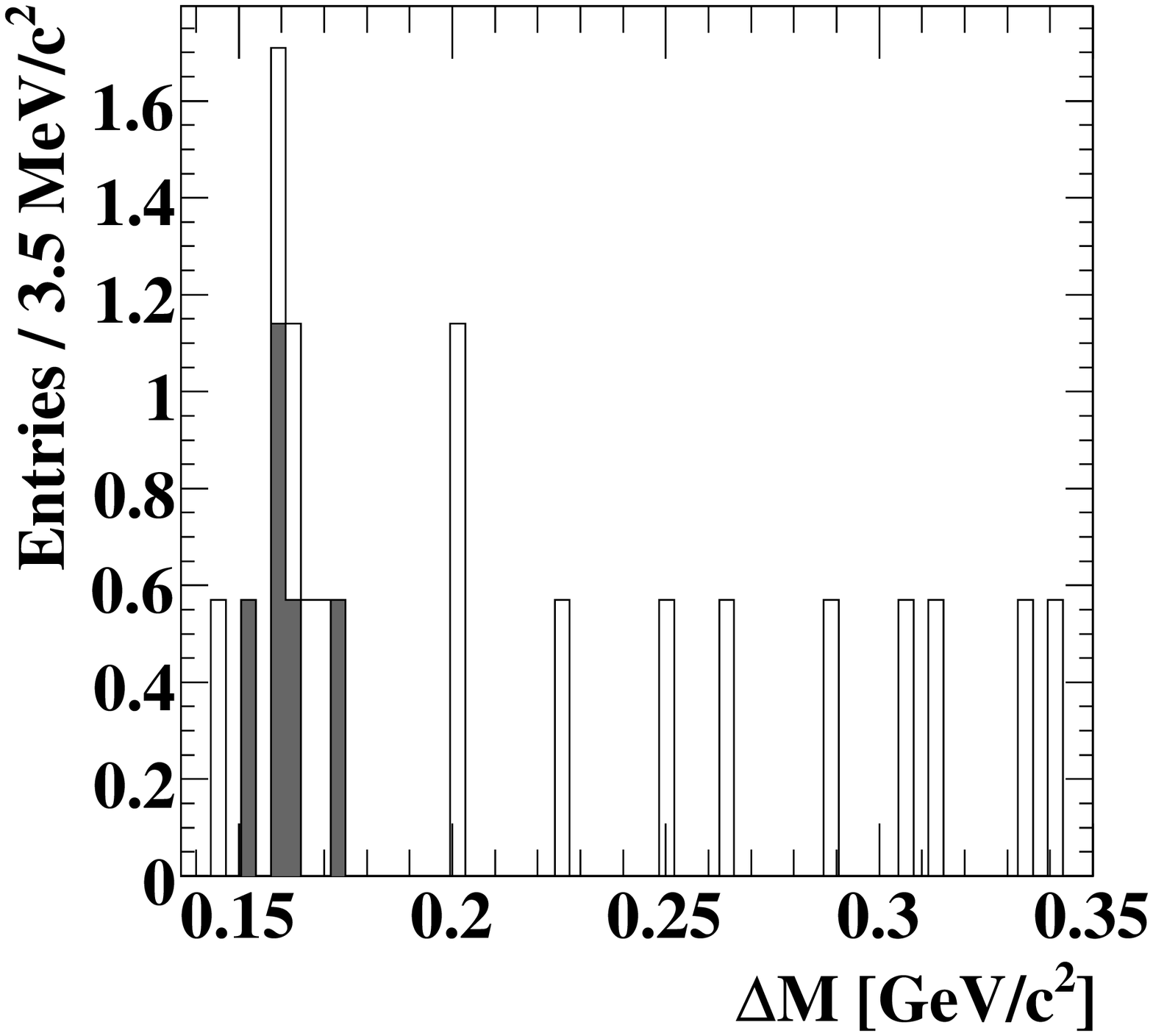}
\caption{WS $\Delta M$ distribution for \babar's
\dklnu\ candidate sample~\cite{babar_semi}, in 
data (left) and MC (right). The MC sample is 
1.75 times the size of the data sample. The 
dark (light) histogram 
shows events after (before) the final kinematic 
selection. In the left plot, region ``1'' denotes 
the signal region used to determine $R^{}_M$.}
\label{fig:semi_babar}
\end{figure}

The final candidate samples for WS data and Monte Carlo (MC) 
simulation are shown in Figs.~\ref{fig:semi_babar}a and 
\ref{fig:semi_babar}b, respectively. In the data, three 
events are observed in the signal region $\Delta M<0.20$\gevm, 
whereas the MC predicts (for the luminosity of the data) 
$2.85\pm 1.43$ background events. Together these values 
give $R^{}_M=(0.004\,^{+0.07}_{-0.06})$\%, where the error 
corresponds to where the log-likelihood function for 
$N^{}_s$ (the true number of signal events) rises by 0.50 
units with respect to the minimum value. The points where 
the log-likelihood function rises by 1.35 units give a 
90\% C.L.\ constraint $-0.13\%\!<\!R^{}_M\!<\!0.12\%$.
This upper bound is similar to the upper limit obtained 
by Belle.

\section{{\boldmath Hadronic Decays $D^0\ra h^+_1\,h^-_2$}}

Two-body hadronic final states $K^+K^-$, $\pi^+\pi^-$, and 
$K^\pm\pi^\mp$ can be reached from either $D^0$ or $\dbar$;
thus $D^0\ra f$ and $D^0\ra\dbar\ra f$ amplitudes both contribute 
to the decay rate, and detecting the effect of the latter provides 
evidence for mixing. The time dependence of the decay rate ${\cal R}$
is given by
\begin{eqnarray}
e^{-\Gamma t}\left[ 1
+ (y\,{\rm Re}\lambda - x\,{\rm Im}\lambda)t
+ |\lambda|^2\,\frac{x^2+y^2}{2}\,t^2\right]\,,
\label{eqn:master}
\end{eqnarray}
where $\lambda=(q/p){\cal A}(\dbar\ra f)/{\cal A}(D^0\ra f)$
and $q,p$ are complex coefficients relating mass eigenstates 
to flavor eigenstates: $D^{}_{1,2}=p|D^0\rangle\pm q|\dbar\rangle$.
The parameter $\lambda$ can be written 
$-|q/p||{\cal A}^{}_{\dbar}/{\cal A}^{}_{D^0}|\,e^{i(\phi+\delta)}$, 
where $\delta$ is the strong phase difference between amplitudes 
${\cal A}(\dbar\ra f)$ and ${\cal A}(D^0\ra f)$, and
$\phi$ is a possible weak phase. In the absence of 
\cp\ violation (\cpv), 
$|q/p|=1$ and $\phi=0$. For $f=K^-\pi^+$, 
${\cal A}(\dbar\ra f)$ is doubly-Cabibbo-suppressed,
${\cal A}(D^0\ra f)$ is Cabibbo-favored, and thus 
$|\lambda|\ll 1$ and ${\cal R}\approx e^{-\Gamma\,t}$.

\subsection{{\boldmath \cp-eigenstates $K^+K^-$ and $\pi^+\pi^-$}}

For decays to self-conjugate states $K^+K^-$ and
$\pi^+\pi^-$, $|\lambda|=|q/p|\approx\!1$ and the
third term in Eq.~(\ref{eqn:master}) can be neglected
since $|x|$ and $|y|$ are very small. As $\delta\!=\!0$, 
if there is no \cpv\ in mixing ($|q/p|\!=\!1$) then
$\lambda=-e^{i\phi}$ and 
${\cal R}^{}_{h^+h^-}\approx e^{-\Gamma t}[1-(y\cos\phi)\Gamma t]
\approx e^{-\Gamma(1+y\cos\phi)\,t}$. Thus
$\tau^{}_{K^-\pi^+}/\tau^{}_{h^+h^-}\!\approx\!1+y\cos\phi$.
The observable
$\tau^{}_{K^-\pi^+}/\tau^{}_{h^+h^-}-1$ is denoted
$y^{}_{CP}$ and, for no \cpv\ in mixing, equals $y\cos\phi$; 
if \cp\ is conserved, $y^{}_{CP}\!=\!y$. Allowing for
arbitrary \cpv, one obtains~\cite{Nir}
\begin{eqnarray*}
y^{}_{CP} \hskip-0.05in & = & \hskip-0.05in 
\frac{1}{2}
\left(\left|\frac{q}{p}\right|+\left|\frac{p}{q}\right|\right)y\cos\phi -
\frac{1}{2}
\left(\left|\frac{q}{p}\right|-\left|\frac{p}{q}\right|\right)x\sin\phi\,.
\end{eqnarray*}
The (normalized) difference in lifetimes 
$A^{}_\Gamma\equiv (\tau^{}_{D^0\rightarrow K^+K^-}\!-
\tau^{}_{\dbar\rightarrow K^+K^-})/\tau^{}_{K^+K^-}$ is 
equal to the related expression~\cite{Nir}
\begin{eqnarray*}
A^{}_\Gamma \hskip-0.05in & = & \hskip-0.05in 
\frac{1}{2}
\left(\left|\frac{q}{p}\right|-\left|\frac{p}{q}\right|\right)y\cos\phi -
\frac{1}{2}
\left(\left|\frac{q}{p}\right|+\left|\frac{p}{q}\right|\right)x\sin\phi\,.
\end{eqnarray*}

This method has been used by numerous experiments to
constrain \ycp~\cite{ycp_references}. Belle's new
measurement~\cite{belle_kk} uses 540~fb$^{-1}$ of data 
and both $K^+K^-$ and $\pi^+\pi^-$ final states. One advance 
of this analysis is the resolution function, which is
constructed as a sum over many Gaussian functions $G$:
\begin{eqnarray*}
{\cal R}(t-t^{}_{\rm true}) & = &
\sum_{i=1}^{n} f^{}_i 
\sum_{k=1}^{3} w^{}_k\,G(t-t^{}_{\rm true};\sigma^{}_{ik},t^{}_0)\,,
\end{eqnarray*}
with standard deviations
$\sigma^{}_{ik}=s^{}_k\times\sigma^{\rm pull}_k\times\sigma^{}_i$.
In this expression, $f^{}_i$ is the weight of the value $\sigma^{}_i$ 
taken from the normalized, binned, $D^0\ra K^-\pi^+$ distribution 
of $\sigma^{}_t$, the event-by-event uncertainty in the decay time 
(i.e., $\sigma^{}_i=\sigma^{}_t({\rm bin\ }i)$). Parameter $w^{}_k$ 
is the weight of value $\sigma^{\rm pull}_k$ obtained by fitting 
the MC pull distribution to a sum of three Gaussians with widths 
$\sigma^{\rm pull}_k\ (k=\!1\!-\!3)$. The $s^{}_k$ are scale factors 
to account for differences between MC and data, and $t^{}_0$ 
is a common offset. The parameters $s^{}_k$ and $t^{}_0$ are 
left free when fitting for~\ycp. This resolution function, 
and a slight variation with an additional offset parameter,
yields accurate values of the $D^0\ra K^-\pi^+$ lifetime over 
all running periods. The mean value is $408.7\,\pm 0.6$~fs, 
which is consistent with the PDG value~\cite{pdg} (and actually 
has greater statistical precision).

\begin{figure}[h]
\centering
\includegraphics[width=80mm]{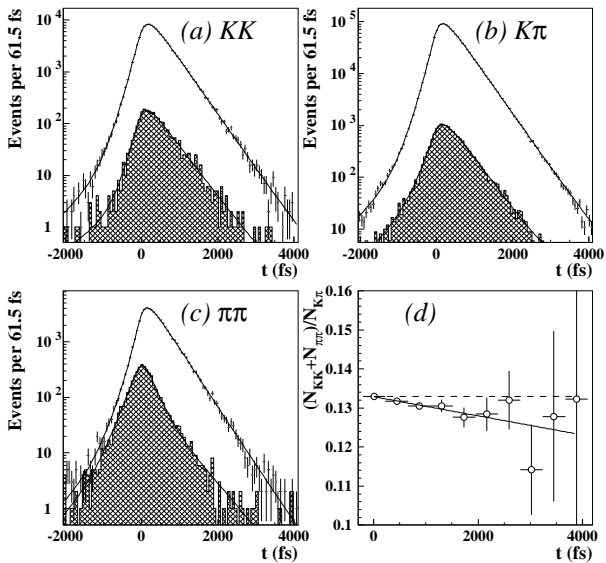}
\caption{Projections of the decay-time fit superimposed 
on the data for \dkk, $D^0\ra K^-\pi^+$, and \dpipi\ decays, 
from Belle~\cite{belle_kk}. The hatched area represents the 
background contribution. Plot {\it (d)} shows the ratio of 
decay-time distributions for $D^0\ra(K^+K^- +\pi^+\pi^-)$ 
and $D^0\ra K^-\pi^+$; the solid line is a fit to the points.}
\label{fig:ycp_belle}
\end{figure}

Fitting the $K^-\pi^+$, $K^+K^-$, and $\pi^+\pi^-$ decay 
time distributions (Figs.~\ref{fig:ycp_belle}a-c) shows a 
statistically significant difference between the $K^-\pi^+$ and 
$h^+h^-$ lifetimes. 
The effect is visible in Fig.~\ref{fig:ycp_belle}d, which plots
the ratio of event yields $N^{}_{h^+h^-}/N^{}_{K\pi}$ as a 
function of decay time. Performing a simultaneous maximum 
likelihood (ML) fit to all three $h^+_1 h^-_2$ samples gives 
\begin{eqnarray}
y^{}_{\rm CP} & = & (1.31\,\pm 0.32\,\pm 0.25)\%\,,
\end{eqnarray}
which deviates from zero by $3.3\sigma$. The systematic
error is dominated by uncertainty in the background decay time 
distribution, variation of selection criteria, and the 
assumption that $t^{}_0$ is equal for all three final 
states. The analysis also measures
\begin{eqnarray}
A^{}_\Gamma & = & (0.01\,\pm 0.30\,\pm 0.15)\%\,,
\end{eqnarray}
which is consistent with zero (no \cpv). The sources of 
systematic error for $A^{}_\Gamma$ are similar to those for~\ycp.

\subsection{\boldmath ``Wrong-sign'' \dkpi\ Decays}

For $f\!=\!K^+\pi^-$ and no \cpv,
Eq.~(\ref{eqn:master}) simplifies to
$e^{-\Gamma t}\left[ R^{}_D +\sqrt{R^{}_D}\,y'\,t + (x'^2+y'^2)\,t^2/4 \right]$,
where 
$R^{}_D={\cal A}(D^0\ra K^+\pi^-)/{\cal A}(\dbar\ra K^+\pi^-)$
and $x'=x\cos\delta + y\sin\delta$, $y'=y\cos\delta - x\sin\delta$ 
are ``rotated'' mixing parameters. Both \babar~\cite{babar_kpi} 
and Belle~\cite{belle_kpi} do unbinned ML fits to the decay-time 
distribution of WS $D^0\ra K^+\pi^-$ decays to determine $x'^2, y'$, 
and $R^{}_D$. Because $K^+\pi^-$ is doubly-Cabibbo-suppressed, 
$R^{}_D$ is small and the mixing terms in the above 
expression play a larger role; however, there is 
substantial background, $\sim$\,48\%. The largest background 
component consists of real $D^0\ra K^-\pi^+$ decays combining 
with random $\pi^-$ tracks; fortunately, the decay time 
distribution for this background is simple, the same as 
that for RS $D^0\ra K^-\pi^+$ decays.

\begin{figure}[h]
\centering
\includegraphics[width=80mm]{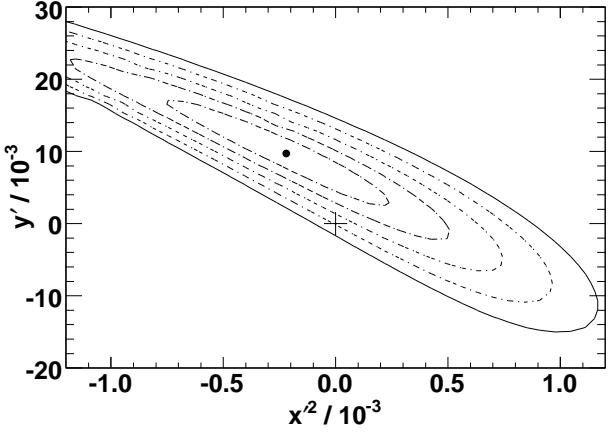}
\caption{Two-dimensional $1\sigma$-$5\sigma$ contours
for $(x'^2,y')$, from \babar's fit to the decay time 
distribution of \dkpi\ decays~\cite{babar_kpi}. The 
contours are calculated from the change in the value 
of $-2\ln{\cal L}$ from the minimum value, and
systematic uncertainties are included.
The point is the best-fit value. }
\label{fig:kpi_babar}
\end{figure}

\begin{figure}[h]
\centering
\includegraphics[width=80mm]{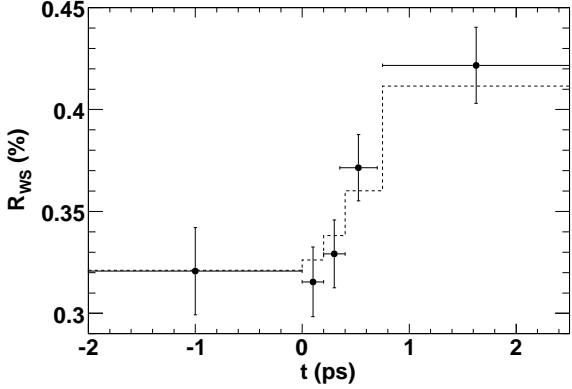}
\caption{
The WS \dkpi\ branching fraction for separate 
bins of decay time, from \babar~\cite{babar_kpi}. 
The dashed line shows the expectation based on 
the values of $(x'^2,y')$ obtained from the decay 
time fit. The data agrees well with the dashed line 
but not with a flat (i.e., no mixing) line. }
\label{fig:kpi_babar_backsubtracted}
\end{figure}

The results of the \babar\ and Belle fits are listed 
in Table~\ref{tab:kpi_results}. For \babar, the value 
for $x'^2$ is negative, i.e., outside the physical region, 
nominally due to statistical fluctuation. The \babar\ likelihood
contours are shown in Fig.~\ref{fig:kpi_babar}. The no-mixing
point $(0,0)$ has $\Delta{\cal L}=23.9$ units above the minimum 
value, corresponding to a CL of only 0.01\% ($3.9\sigma$) 
including systematic uncertainty. This constitutes 
evidence for mixing. The largest systematic error is 
from uncertainty in modeling the tail of the background
decay time distribution. The mixing is visible in 
Fig.~\ref{fig:kpi_babar_backsubtracted}, which plots
the ratio of the background-subtracted yields of 
WS to RS decays in bins of decay time. For each bin, 
the yields are determined from two-dimensional fits 
to variables $M(K\pi)$ and $\Delta M\equiv M(\pi^{}_s\,K\pi)-M(K\pi)$.
The plot shows the ratio increasing with decay time, consistent 
with Eq.~(\ref{eqn:master}) but inconsistent 
with the no-mixing or flat hypothesis.
Fitting to Eq.~(\ref{eqn:master}) gives
$\chi^2/{\rm dof}\!=\!1.5$, whereas fitting to a flat
distribution gives $\chi^2/{\rm dof}\!=\!24.0$.
To allow for \cpv, \babar\ fits the $D^0$ and $\dbar$ 
samples separately; the results are consistent with each 
other, showing no evidence of \cpv\ (see Table~\ref{tab:kpi_results}). 

\begin{table}[h]
\begin{center}
\caption{\babar~\cite{babar_kpi} and Belle~\cite{belle_kpi} results 
from fitting the decay time distribution of $D^0\ra K^+\pi^-$ decays. 
The errors listed are statistical plus systematic, except for those 
from \babar's $D^0$ and $\dbar$ subsamples, which are statistical only.}
\label{tab:kpi_results}
\begin{tabular}{|l|c|c|c|}
\hline 
\textbf{Exp.\,(fb$^{-1}$)} & \textbf{$x'^2$ (\%)} 
& \textbf{$y'$ (\%)} & \textbf{$R^{}_D$ (\%)} \\
\hline 
\babar\ (384) & $-0.022\,\pm 0.037$ & $0.97\,\pm 0.54$ & $0.303\,\pm 0.019$ \\
\ \ $D^0$ only   & $-0.024\,\pm 0.043$ & $0.98\,\pm 0.64$ &  \\
\ \ $\dbar$ only & $-0.020\,\pm 0.041$ & $0.96\,\pm 0.61$ &  \\
\hline
Belle\ (400) & $0.018\,^{+0.021}_{-0.023}$ 
& $0.06\,^{+0.40}_{-0.39}$ & $0.364\,\pm 0.017$ \\
\ \cpv-allwd & $< 0.072$ & $(-2.8, 2.1)$ &  \\
\hline
\end{tabular}
\end{center}
\end{table}

\begin{figure}[h]
\centering
\includegraphics[width=80mm]{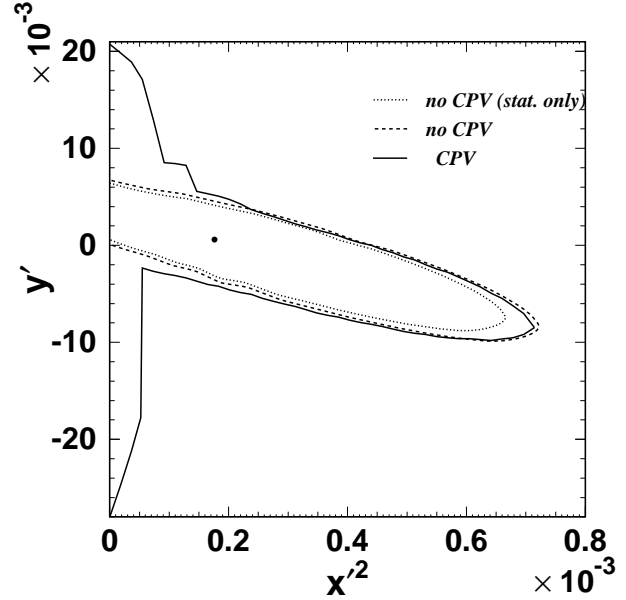}
\caption{95\% C.L. contours for $(x'^2,y')$ from 
Belle's analysis of \dkpi\ decays~\cite{belle_kpi}: 
dotted (dashed) is statistical (statistical plus 
systematic) contour for no \cpv; solid is statistical 
plus systematic contour allowing for \cpv. The point 
is the best-fit value for no~\cpv. The contours are 
obtained from a frequentist toy-MC calculation. }
\label{fig:kpi_belle}
\end{figure}

The Belle measurement has somewhat greater statistical
precision than that of \babar, but the central values
are in the physical region ($x'^2>0$). Belle obtains 
confidence regions for $x'^2$ and $y'$ using a toy-MC 
frequentist method. Due to the proximity of the unphysical 
region, the procedure uses Feldman-Cousins likelihood ratio 
ordering~\cite{FeldmanCousins}. The resulting contours are 
shown in Fig.~\ref{fig:kpi_belle}. The CL of the no-mixing 
point $(0,0)$ is 3.9\%, corresponding to $2.1\sigma$. The 
largest systematic uncertainty arises from variation of 
the $p^{}_{D^*}$ minimum value cut.

Belle searches for \cpv\ by fitting the $D^0$ and $\dbar$ 
distributions separately. The results, denoted 
$(x'^{2+},y'^+,R^{}_D)$ and $(x'^{2-},y'^-,\overline{R}^{}_D)$,
respectively, are used to calculate the
\cpv\ parameter $A^{}_M=(R^+_M-R^-_M)/(R^+_M+R^-_M)$,
where $R^\pm_M=(x'^{\pm 2}+y'^{\pm 2})/2$. Theoretically,
$A^{}_M=(|q/p|^4-1)/(|q/p|^4+1)\approx |q/p|^2-1$. The 
\cpv\ parameter ${\rm Arg}(q/p)\equiv\phi$ and mixing
parameters $x'^2,\,y'$ are determined via the relations
\begin{eqnarray*}
x'^\pm &= &[(1\pm A^{}_M)/(1\mp A^{}_M)]^{1/4}(x'\cos\phi\pm y'\sin\phi) \\
y'^\pm &= &[(1\pm A^{}_M)/(1\mp A^{}_M)]^{1/4}(y'\cos\phi\mp x'\sin\phi)\,.
\end{eqnarray*}
The resulting confidence region for $(x'^2,y')$
is plotted in Fig.~\ref{fig:kpi_belle} as the solid 
contour; the complicated shape is due to there being two 
solutions for $(x',y')$, depending on the relative sign of 
$x'^+$ and~$x'^-$ (which is unmeasured). The fit results 
are $\phi=(9.4\,\pm 25.3)^\circ$ or $(84.5\,\pm 25.3)^\circ$ 
for the same or opposite signs of $x'^+$ and $x'^-$,
and $A^{}_M=0.67\,\pm 1.20$.

\section{\boldmath Dalitz Plot Analysis of \dkspp}

The time dependence of the Dalitz plot for \dkspp\ decays 
is sensitive to mixing parameters $x$ and $y$ without ambiguity 
due to strong phases.
For a particular point in the Dalitz plot $(m^2_+,m^2_-)$, where 
$m^{}_+\equiv m(K^0_S\,\pi^+)$ and $m^{}_-\equiv m(K^0_S\,\pi^-)$, 
the overall decay amplitude is
\begin{eqnarray}
{\cal A}^{}_{D^0}(m^2_+,m^2_-)\,\frac{e^{}_1(t)+e^{}_2(t)}{2} & + & \nonumber \\
 & & \hskip-1.2in 
\left(\frac{q}{p}\right)
{\cal A}^{}_{\dbar}(m^2_+,m^2_-)\,\frac{e^{}_1(t)-e^{}_2(t)}{2}\,,
\label{eqn:kspp_amplitudes}
\end{eqnarray}
where $e^{}_{(1,2)}(t)=e^{-(im^{}_{1,2}+\Gamma^{}_{1,2}/2)t}$.
The first term represents the (time-dependent) amplitude for 
\dkspp, and the second term represents the amplitude 
for $D^0\!\ra\!\dbar\!\ra\!K^0_S\,\pi^+\pi^-$. Taking the 
modulus squared of Eq.~(\ref{eqn:kspp_amplitudes}) gives 
the decay rate or, equivalently, the density of points 
$\rho(m^2_+,m^2_-;t)$. The result contains terms proportional 
to  $\cosh(y\,\Gamma t)$, $\cos(x\,\Gamma t)$, and 
$\sin(x\,\Gamma t)$, and thus fitting the time-dependence 
of $\rho(m^2_+,m^2_-;t)$ determines $x$ and~$y$.
This method was developed by CLEO~\cite{cleo_kspp}.

To use Eq.~(\ref{eqn:kspp_amplitudes}) requires 
choosing a model for the decay amplitudes 
${\cal A}^{}_{D^0,\dbar}(m^2_+,m^2_-)$. This is usually taken 
to be the ``isobar model''~\cite{isobar}, and thus, in addition 
to $x$ and $y$, one also fits for the magnitudes and phases of 
various intermediate states. Specifically,
${\cal A}^{}_{D^0}(m^2_+,m^2_-)=\sum_j a^{}_j\,e^{i\delta^{}_j}\,A^{}_j$,
where $\delta^{}_j$ is a strong phase, $A^{}_j$ is the
product of a relativistic Breit-Wigner function and
Blatt-Weiskopf form factors, and the parameter $j$
runs over all intermediate states. This sum includes
possible scalar resonances and, typically, a constant 
non-resonant term. For no direct \cpv, 
${\cal A}^{}_{\dbar}(m^2_+,m^2_-)={\cal A}^{}_{D^0}(m^2_-,m^2_+)$;
otherwise, one must consider separate decay parameters
$(a^{}_j,\delta^{}_j)$ for $D^0$ decays and
$(\bar{a}^{}_j,\bar{\delta}^{}_j)$ for $\dbar$ 
decays. 

Belle has recently fit a large \dkspp\ sample selected
from 540~fb$^{-1}$ of data~\cite{belle_kspp}. The analysis 
proceeds in two steps. First, signal and background yields are 
determined from a two-dimensional fit to variables $M(K\pi\pi)$ 
and $\Delta M = M(\pi^{}_s K\pi\pi)-M(K\pi\pi)$. Within a signal 
region $|M(K\pi\pi)-M^{}_{D^0}|\!<\!15$\mevm\ and
$|\Delta M-5.9{\rm\ MeV}|\!<\!1.0$~MeV (corresponding 
to $3\sigma$ in resolution), there are 534\,000 signal 
candidates with 95\% purity. These events are fit for 
$x$ and $y$; the (unbinned ML) fit variables are
$m^2_+$, $m^2_-$, and the decay time~$t$. Most of the 
background is combinatoric, i.e., the $D^0$ candidate 
results from a random combination of tracks. The decay-time 
distribution of this background is modeled as the sum of 
a delta function and an exponential function convolved 
with a Gaussian resolution function, and all parameters 
are determined from fitting events in the sideband
$30{\rm\ MeV}/c^2<|M(K\pi\pi)-M^{}_{D^0}|<55$\mevm.

\begin{figure}
\centering
\includegraphics[width=39mm]{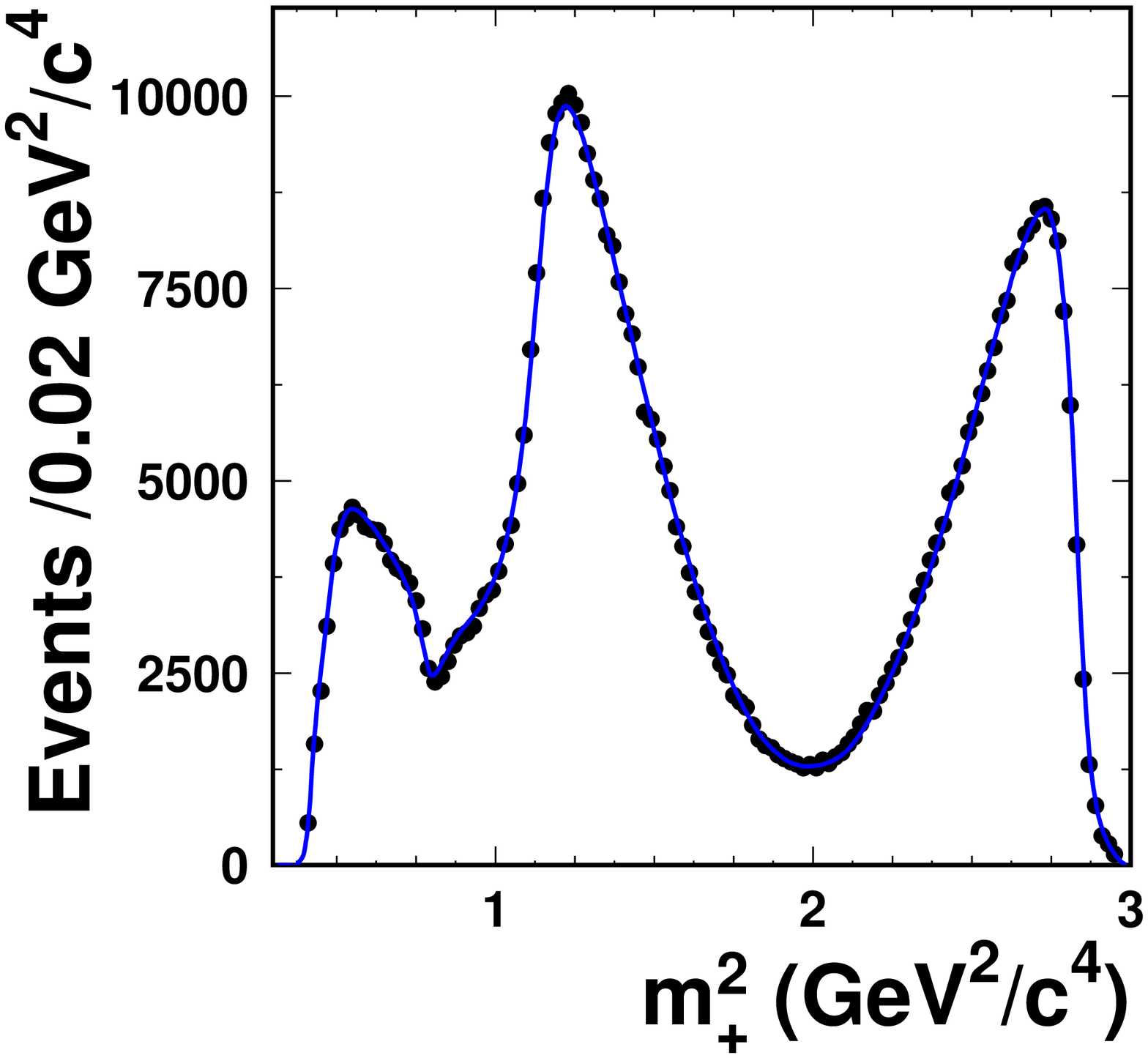}
\hskip0.05in
\includegraphics[width=39mm]{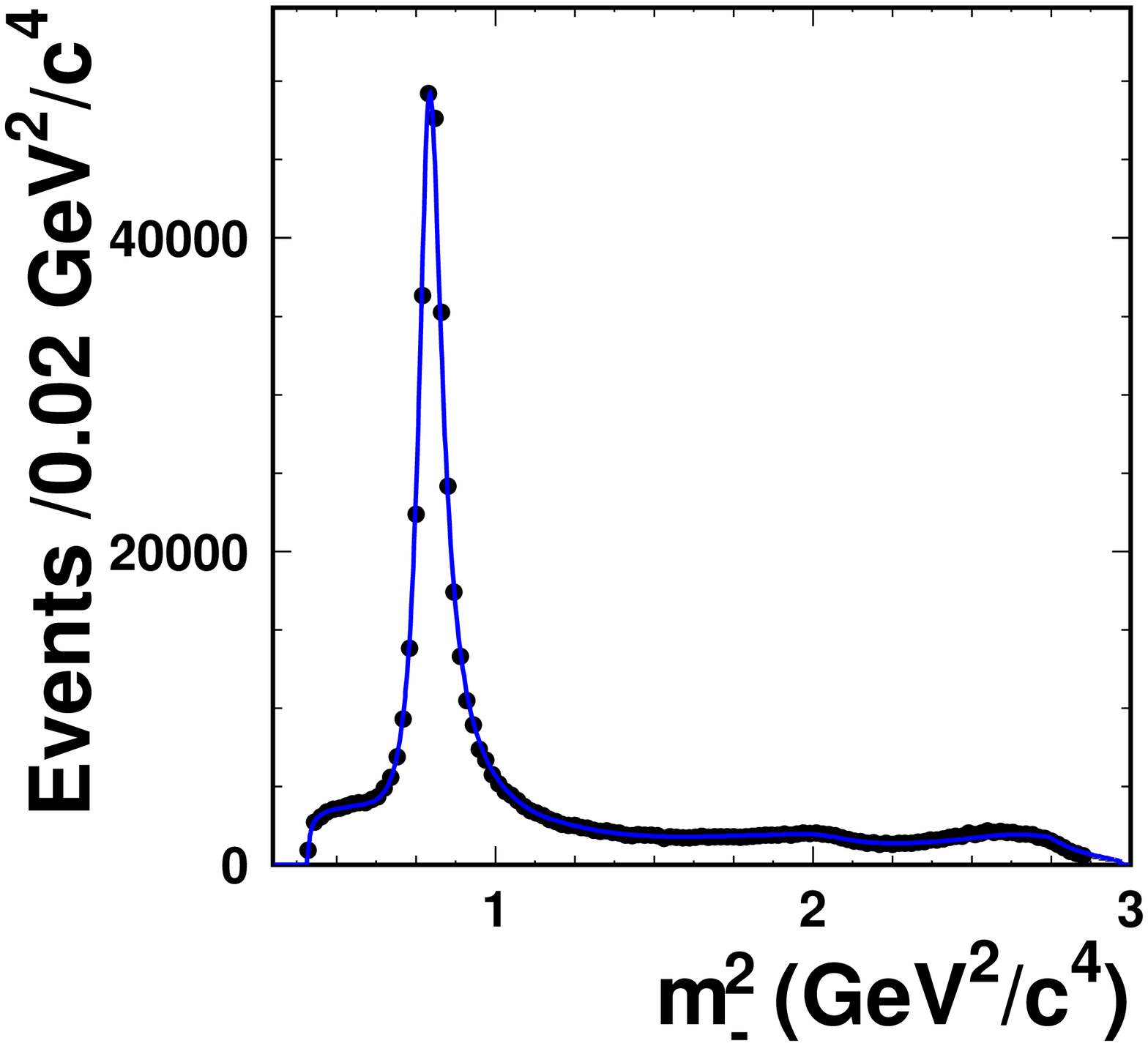}
\vskip0.10in
\includegraphics[width=39mm]{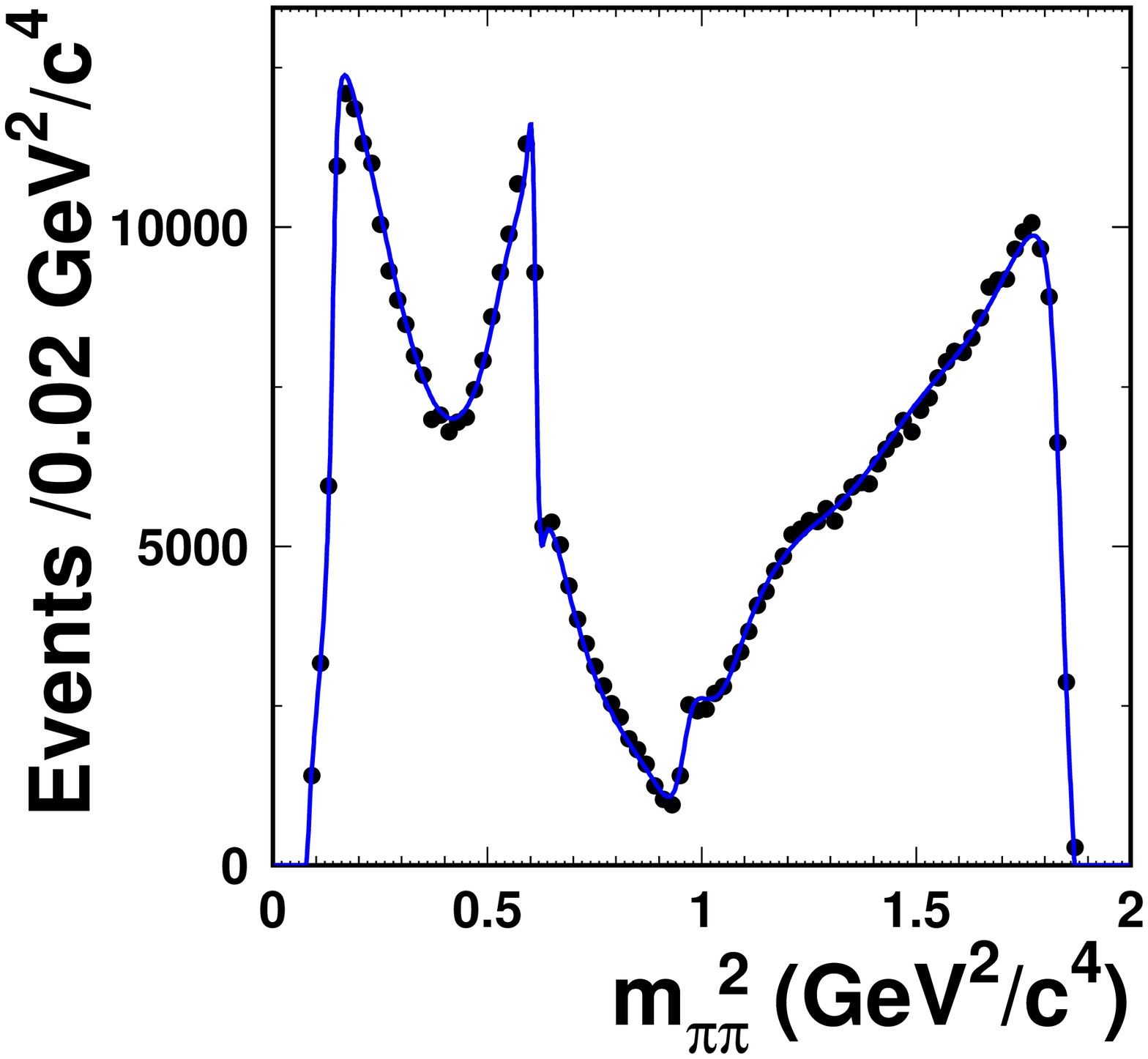}
\hskip0.05in
\includegraphics[width=39mm]{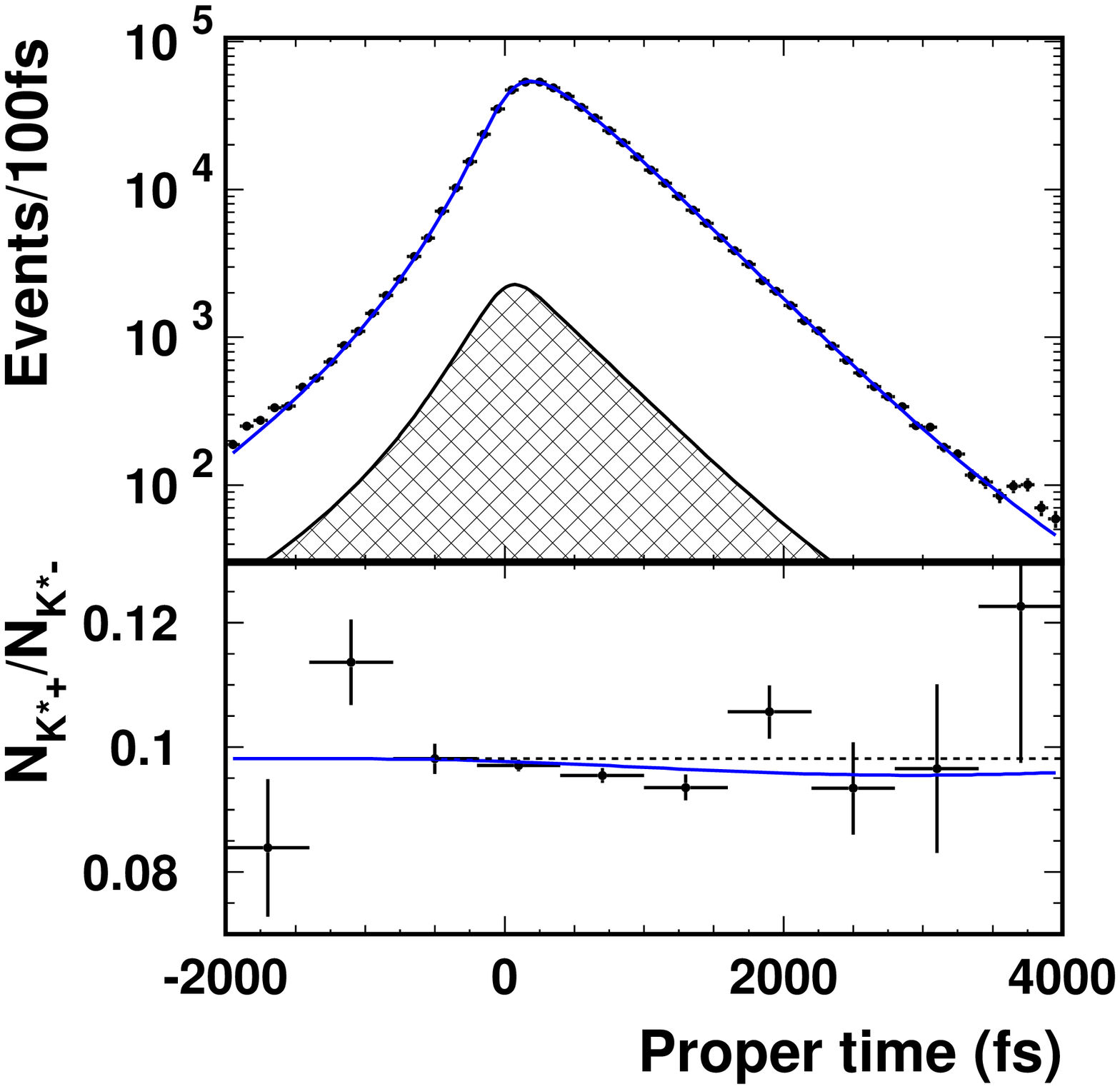}
\caption{Projection of the unbinned ML fit superimposed
on the data for \dkspp\ decays, from Belle~\cite{belle_kspp}. 
In {\it (d)}, the hatched area represents the combinatorial 
background contribution, and the lower plot shows the ratio 
of decay-time distributions for events in the $K^*(892)^+$ 
and $K^*(892)^-$ regions, where sensitivity to $(x,y)$ 
is highest.}
\label{fig:kspp_projections}
\end{figure}

The results from two separate fits are listed in 
Table~\ref{tab:kspp_belle}. In the first fit \cp\ 
conservation is assumed, i.e., $q/p\!=\!1$ and 
${\cal A}^{}_{\dbar}(m^2_+,m^2_-)={\cal A}^{}_{D^0}(m^2_-,m^2_+)$.
The free parameters are $x,y,\tau^{}_{D^0}$, some
timing resolution function parameters, and decay 
model parameters $(a^{}_r,\delta^{}_r)$. The results 
for the latter are listed in Table~\ref{tab:kspp_resonances}. 
The results for $x$ and $y$ indicate that $x$ is
positive, about $2\sigma$ from zero. Projections 
of the fit are shown in Fig.~\ref{fig:kspp_projections}.
The fit also yields $\tau^{}_D=(409.9\pm 0.9)$~fs, 
which is consistent with the PDG value~\cite{pdg} 
(and actually has greater statistical precision).

For the second fit, \cpv\ is allowed and the $D^0$
and $\dbar$ samples are considered separately. This introduces
additional parameters $|q/p|$, ${\rm Arg}(q/p)=\phi$, and
$(\bar{a}^{}_j,\bar{\delta}^{}_j)$. The fit gives two
equivalent solutions, $(x,y,\phi)$ and $(-x,-y,\phi+\pi)$.
Aside from this possible sign change, the effect upon $x$ and 
$y$ is small, and the results for $|q/p|$ and $\phi$ are consistent 
with no~\cpv. The sets of Dalitz parameters $(a^{}_r,\delta^{}_r)$ 
and $(\bar{a}^{}_r,\bar{\delta}^{}_r)$ are consistent with 
each other, indicating no direct~\cpv. Taking
$a^{}_j\!=\!\bar{a}^{}_j$ and $\delta^{}_j\!=\!\bar{\delta}^{}_j$
(i.e., no direct \cpv) and repeating the fit gives 
$|q/p|=0.95\,^{+0.22}_{-0.20}$ and
$\phi = (-2\,^{+10}_{-11})^\circ$.

The dominant systematic errors are from the time 
dependence of the Dalitz plot background, and the 
effect of the $p^{}_{D^*}$ momentum cut used to reject 
$D^*$'s originating from $B$ decays. The default fit 
includes $\pi\pi$ scalar resonances $\sigma^{}_1$ and $\sigma^{}_2$; 
when evaluating systematic errors, the fit is repeated
without any $\pi\pi$ scalar resonances using $K$-matrix 
formalism~\cite{K-matrix}. The influence upon $x$ 
and $y$ is small and included as a systematic error.

\begin{figure}
\centering
\includegraphics[width=80mm]{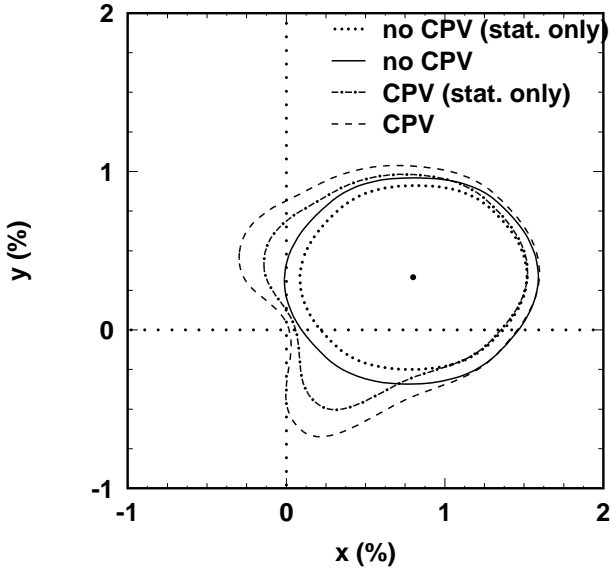}
\caption{95\% C.L. contours for $(x,y)$ from Belle~\cite{belle_kspp}: 
dotted (solid) is statistical (statistical plus systematic) contour
for no \cpv; dashed-dotted (dashed) is statistical
(statistical plus systematic) contour allowing for \cpv.
The point is the best-fit value for no~\cpv.}
\label{fig:kspp_conf_regions}
\end{figure}

The 95\% C.L.\ contour for $(x,y)$ is plotted in 
Fig.~\ref{fig:kspp_conf_regions}. The contour is
obtained from the locus of points where $-2\ln{\cal L}$
rises by 5.99 units from the minimum value; the distance 
of the points from the origin is subsequently rescaled 
to include systematic uncertainty.
We note that for the \cpv-allowed case, the reflections of
the contours through the origin are also allowed regions.

\begin{table}
\renewcommand{\arraystretch}{1.3}
\caption{Fit results and 95\% C.L. intervals for $x$ and $y$,
from Belle's analysis of \dkspp\ decays~\cite{belle_kspp}. The 
errors are statistical, experimental systematic, and decay-model 
systematic, respectively.}
\label{tab:kspp_belle} 
\begin{tabular}{llcc}\hline\hline
Fit   & Param. & Result  & 95\% C.L. inter. \\
\hline
No    & $x\,(\%)$ 
  & $0.80\,\pm 0.29\,^{+0.09\,+0.10}_{-0.07\,-0.14}$ & $(0.0,1.6)$ \\
$CPV$ & $y\,(\%)$ 
  & $0.33\,\pm 0.24\,^{+0.08\,+0.06}_{-0.12\,-0.08}$ & $(-0.34,0.96)$ \\
\hline
$CPV$ & $x\,(\%)$ 
  & $0.81\,\pm0.30\,^{+0.10\,+0.09}_{-0.07\,-0.16}$ & $|x|<1.6$  \\
      & $y\,(\%)$ 
  & $0.37\,\pm0.25\,^{+0.07\,+0.07}_{-0.13\,-0.08}$ & $|y|<1.04$ \\
      & $|q/p|$ &$0.86\,^{+0.30\,+0.06}_{-0.29\,-0.03}\,\pm 0.08$ & $-$ \\
      & $\phi\,(^\circ)$ & $-14\,^{+16\,+5\,+2}_{-18\,-3\,-4}$ & $-$ \\
\hline\hline
\end{tabular}
\end{table}

\begin{table}
\caption{Fit results for \dkspp\ Dalitz plot parameters,
from Belle~\cite{belle_kspp}. The errors are statistical only. 
The fit fraction is defined as the ratio of the integral
$\int |a^{}_r\,{\cal A}^{}_r(m^2_-\,,m^2_+)|^2\,dm^2_-\,dm^2_+$ 
to 
$\int |\sum_{r=1}^n a^{}_r\,e^{i\phi^{}_r}
     {\cal A}^{}_r(m^2_-\,,m^2_+)|^2\,dm^2_-\,dm^2_+$.}
\label{tab:kspp_resonances}
\begin{tabular}{lccc}\hline\hline
Resonance & Amplitude& Phase (deg) &Fit fraction \\
\hline
$K^*(892)^-$    & $1.629\pm0.006$  & $134.3\pm0.3$  & 0.6227 \\
$K_0^*(1430)^-$ & $2.12\pm0.02$    & $-0.9\pm0.8$   & 0.0724 \\
$K_2^*(1430)^-$ & $0.87\pm0.02$    & $-47.3\pm1.2$  & 0.0133 \\
$K^*(1410)^-$   & $0.65\pm0.03$    & $111\pm4$      & 0.0048 \\
$K^*(1680)^-$   & $0.60\pm0.25$    & $147\pm29 $     & 0.0002 \\
\hline
$K^*(892)^+$    & $0.152\pm0.003$  & $-37.5\pm1.3$  & 0.0054 \\
$K_0^*(1430)^+$ & $0.541\pm0.019$  & $ 91.8\pm2.1$  & 0.0047 \\
$K_2^*(1430)^+$ & $0.276\pm0.013$  & $-106\pm3$     & 0.0013 \\
$K^*(1410)^+$   & $0.33\pm0.02$  & $-102\pm4$     & 0.0013 \\
$K^*(1680)^+$   & $0.73\pm0.16$    & $103\pm11$      & 0.0004 \\
\hline
$\rho(770)$     & 1 (fixed)          & 0 (fixed)      & 0.2111 \\
$\omega(782)$   & $0.0380\pm0.0007$  & $115.1\pm1.1$  & 0.0063 \\
$f_0(980)$      & $0.380\pm0.004$    & $-147.1\pm1.1$ & 0.0452 \\
$f_0(1370)$     & $1.46\pm0.05$      & $98.6 \pm1.8$  & 0.0162 \\
$f_2(1270)$     & $1.43\pm0.02$      & $-13.6\pm1.2$  & 0.0180 \\
$\rho(1450)$    & $0.72\pm0.04$      & $41\pm7$   & 0.0024 \\
$\sigma_1$      & $1.39\pm0.02$    & $-146.6\pm0.9$     & 0.0914 \\
$\sigma_2$      & $0.267\pm0.013$    & $-157\pm3 $    & 0.0088 \\ 
\hline
 NR             & $2.36\pm0.07$      & $155\pm2$      & 0.0615 \\ \hline\hline
\end{tabular}
\end{table}

\section{\boldmath Combining all Measurements}

All mixing measurements can be combined to obtain
world average (WA) values for $x$ and $y$. The Heavy Flavor
Averaging Group (HFAG) has done such a combination by 
adding together log-likelihood functions obtained from 
analyses of $D^0\ra K^+\ell\nu$, $D^0\ra h^+h^-$, $D^0\ra K^+\pi^-$,
$D^0\ra K^+\pi^-\pi^0$, $D^0\ra K^+\pi^-\pi^+\pi^-$, and
\dkspp\ decays, as well as CLEOc results for double-tagged 
branching fractions measured at the $\psi(3770)$ 
resonance~\cite{hfag_charm}. 
The combination of likelihood functions preserves 
correlations among parameters and also accounts for 
non-Gaussian errors. When using this method, HFAG 
assumes negligible~\cpv.

\begin{figure}
\centering
\includegraphics[width=80mm]{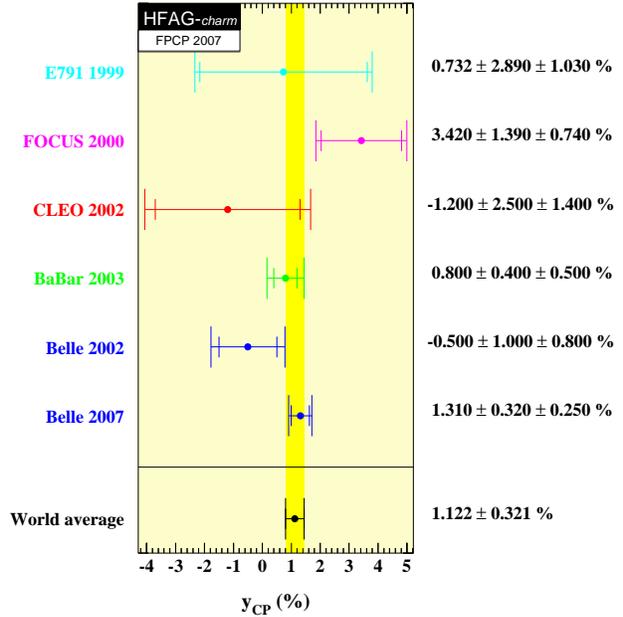}
\caption{World average value for \ycp, from HFAG~\cite{hfag_charm}.}
\label{fig:hfag_ycp}
\end{figure}

\begin{figure}
\centering
\includegraphics[width=80mm]{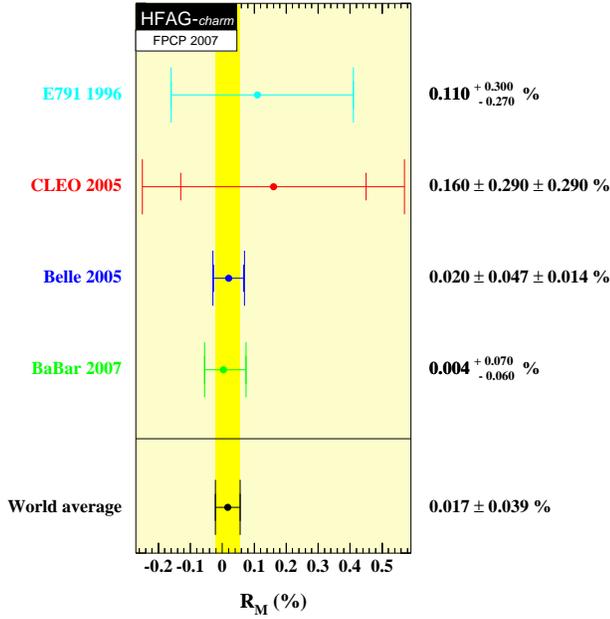}
\caption{World average value for $R^{}_M$, from HFAG~\cite{hfag_charm}.}
\label{fig:hfag_rm}
\end{figure}

As a first step, WA values for \ycp\ and $R^{}_M$ are 
calculated by taking weighted averages of independent 
experimental measurements -- see Figs.~\ref{fig:hfag_ycp}
and \ref{fig:hfag_rm}.These results are then converted to 
three-dimensional likelihood functions for $(x,y,\delta)$. 
For example, the measurement of $y^{}_{CP}$ gives a parabolic
log-likelihood function in $y$ and flat distributions in
$x$ and $\delta$. The $R^{}_M$ likelihood function is an 
annulus in the $x$-$y$ plane and a flat distribution in $\delta$. 

\begin{figure}
\centering
\vskip0.30in
\includegraphics[width=80mm]{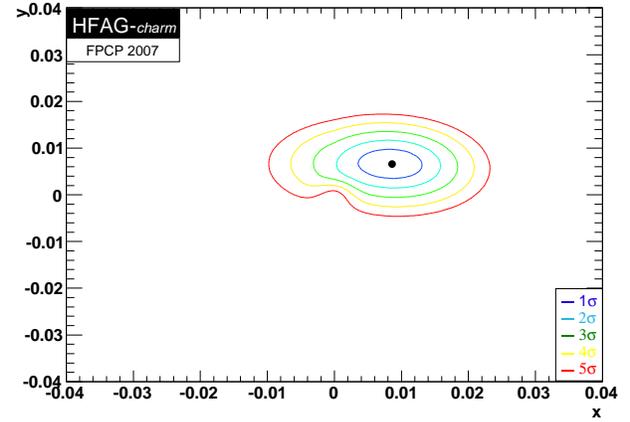}
\caption{Two-dimensional $1\sigma$-$5\sigma$ contours
for $(x,y)$, obtained by adding log-likelihoods from 
measurements of $D^0\ra K^+\ell\nu$, $D^0\ra h^+h^-$, 
$D^0\ra K^+\pi^-$, $D^0\ra K^+\pi^-\pi^0$, 
$D^0\ra K^+\pi^-\pi^+\pi^-$, and \dkspp\ decays, 
and double-tagged branching fractions measured 
at the $\psi(3770)$ resonance (from HFAG~\cite{hfag_charm}).}
\label{fig:hfag_contours}
\end{figure}

The logarithm of the likelihood functions are added, and the 
result is added to the $(x,y,\delta)$ log-likelihood function 
obtained from $D^0\ra K^+\pi^-$ decays. The latter is determined 
as follows. The experiments directly measure a likelihood function
${\cal L}(x'^2,y',R^{}_D)$; thus one first projects out $R^{}_D$ by 
allowing it to take, for any $(x'^2,y')$ point, its preferred 
value. The resulting likelihood for $(x'^2,y')$ is converted
to ${\cal L}(x,y,\delta)$ by scanning values of $(x,y,\delta)$, 
calculating the corresponding values of $(x'^2,y')$, and 
assigning the likelihood for that $(x'^2,y')$ bin. This 
method ignores unphysical (negative) values
of~$x'^2$. The resulting function ${\cal L}(x,y,\delta)$ 
is added to those obtained from \ycp, $R^{}_M$, and other 
measurements. The final likelihood function is projected onto 
the $(x,y)$ plane by letting $\delta$ take, for any $(x,y)$ 
point, its preferred value. This projection is shown in 
Fig.~\ref{fig:hfag_contours}. The unusual shape around 
$x\!=\!y\!=\!0$ is mainly due to $D^0\ra K^+\pi^-$ decays, which 
disfavor the no-mixing point. At $x\!=\!y\!=\!0$, $-2\ln{\cal L}$ 
rises by 37 units above the minimum value; this difference
implies that the no-mixing point is excluded at the level 
of~$5.7\sigma$.

The likelihood function is condensed to one dimension by
letting, for any value of $x$ $(y)$, the parameter $y$ $(x)$
take its preferred value. The resulting likelihoods for 
$(x,y)$ give central values and 68.3\% C.L. intervals
\begin{eqnarray}
x & = & (0.87\,^{+0.30}_{-0.34})\% \\
y & = & (0.66\,^{+0.21}_{-0.20})\%\,.
\end{eqnarray} 
The former is $2.6\sigma$ from zero, and the latter is 
$3.2\sigma$ from zero.

In summary, we conclude the following:
\begin{itemize}
\item the experimental data consistently indicates
that $D^0$'s undergo mixing. The effect is presumably dominated 
by long-distance processes, and unless $|x|\gg |y|$, it may
be difficult to identify new physics from mixing alone.
\item Since \ycp\ is positive, the \cp-even state is shorter-lived,
as in the $K^0$-$\kbar$ system. However, since $x$ appears to be 
positive, the \cp-even state is heavier, unlike in the $K^0$-$\kbar$ system.
\item There is no evidence yet for \cpv\ in the $D^0$-$\dbar$ system.
\end{itemize}

\begin{acknowledgments}
The author thanks the conference organizers for 
excellent hospitality in a beautiful location, and for
assembling a stimulating scientific program.

\end{acknowledgments}


\begin{thebibliography}{99}
\bibitem{charge_conjugates} 
  Charge-conjugate modes are included unless noted otherwise.
\end{thebibliography}
\end{document}